\documentclass[runningheads,a4paper]{llncs}

\usepackage{amssymb}
\usepackage{graphicx}

\usepackage[firstpage]{draftwatermark}
\SetWatermarkText{\hspace*{4.5in}\raisebox{6.3in}{\includegraphics[scale=0.1]{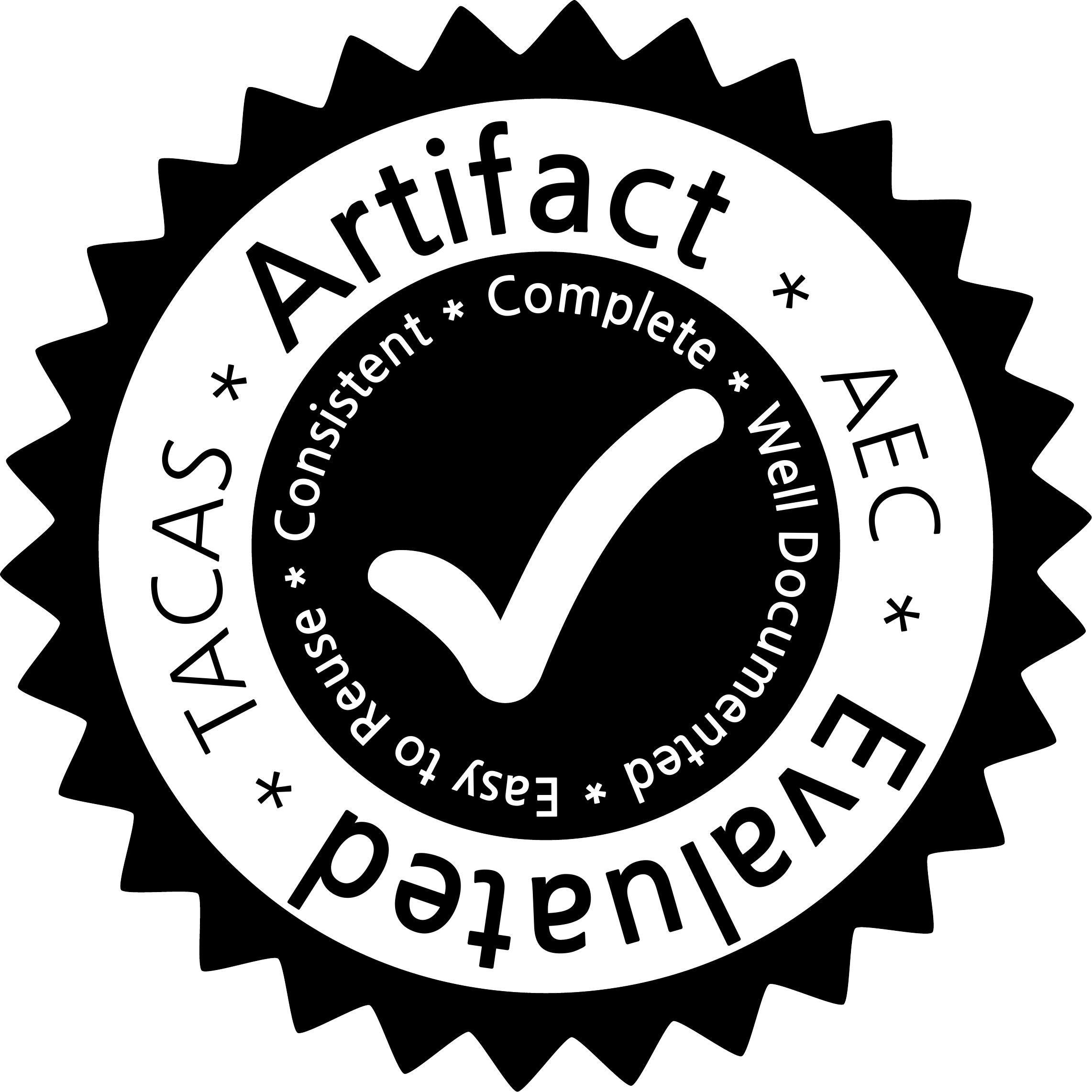}}}
\SetWatermarkAngle{0}

\usepackage{url}
\urldef{\mailsa}\path|{mattarei,barrett}@cs.stanford.edu|    
\urldef{\mailsc}\path|shu@rfrn.org|    
\urldef{\mailsd}\path|{bradnelson,binji}@google.com|    
\urldef{\mailse}\path|jfbastien@apple.com|

\usepackage{todonotes}
\usepackage{amssymb}

\usepackage{amsthm}
\usepackage{xspace}
\usepackage{array}
\usepackage{xargs}
\usepackage{xcolor}
\usepackage{chngcntr}
\usepackage{hyperref}
\usepackage{thmtools}
\usepackage{listings}
\usepackage{subfigure}
\usepackage{paralist}
\usepackage{amsmath,amsfonts}
\usepackage{wrapfig}
\usepackage[ruled,linesnumbered]{algorithm2e}

\usepackage{hyphenat}
\theoremstyle{definition}

\counterwithout{figure}{section}

\makeatletter
\g@addto@macro\normalsize{  \setlength\abovedisplayskip{03pt}
  \setlength\belowdisplayskip{03pt}
  \setlength\abovedisplayshortskip{03pt}
  \setlength\belowdisplayshortskip{03pt}
}
\makeatother

\newcolumntype{L}[1]{>{\raggedright\let\newline\\\arraybackslash\hspace{0pt}}m{#1}}
\newcolumntype{C}[1]{>{\centering\let\newline\\\arraybackslash\hspace{0pt}}m{#1}}
\newcolumntype{R}[1]{>{\raggedleft\let\newline\\\arraybackslash\hspace{0pt}}m{#1}}

\makeatletter
\AtBeginDocument{  \let\c@figure\c@lstlisting
  \let\thefigure\thelstlisting
  \let\ftype@lstlisting\ftype@figure }
\makeatother

\newcommand{\memorymodel}{Memory Model\xspace}
\newcommand{\JS}{JavaScript\xspace}
\newcommand{\ES}{ECMAScript\xspace}
\newcommand{\RBF}{\emph{Reads Bytes From}\xspace}
\newcommand{\HB}{\emph{Happens Before}\xspace}
\newcommand{\MO}{\emph{Memory Order}\xspace}
\newcommand{\SW}{\emph{Synchronizes With}\xspace}
\newcommand{\RF}{\emph{Reads From}\xspace}
\newcommand{\AO}{\emph{Agent Order}\xspace}

\newcommand{\ITE}{\emph{if-then-else}\xspace}
\newcommand{\FOR}{\emph{for-loop}\xspace}

\newcommandx{\unsure}[2][1=]{\todo[linecolor=red,backgroundcolor=red!25,bordercolor=red,#1]{#2}}
\newcommandx{\change}[2][1=]{\todo[inline,linecolor=blue,backgroundcolor=blue!25,bordercolor=blue,#1]{#2}}
\newcommandx{\info}[2][1=]{\todo[linecolor=OliveGreen,backgroundcolor=OliveGreen!25,bordercolor=OliveGreen,#1]{#2}}
\newcommandx{\improvement}[2][1=]{\todo[linecolor=Plum,backgroundcolor=Plum!25,bordercolor=Plum,#1]{#2}}

\newcommandx{\cm}[2][1=]{\todo[inline,linecolor=blue,backgroundcolor=blue!25,bordercolor=black,#1]{\textbf{CM:} #2}}
\newcommandx{\cb}[2][1=]{\todo[inline,linecolor=blue,backgroundcolor=orange!50,bordercolor=black,#1]{\textbf{CB:} #2}}
\newcommandx{\sg}[2][1=]{\todo[inline,linecolor=blue,backgroundcolor=red!25,bordercolor=black,#1]{\textbf{SG:} #2}}
\newcommandx{\bn}[2][1=]{\todo[inline,linecolor=blue,backgroundcolor=green!25,bordercolor=black,#1]{\textbf{BN:} #2}}
\newcommandx{\bs}[2][1=]{\todo[inline,linecolor=blue,backgroundcolor=yellow!25,bordercolor=black,#1]{\textbf{BS:} #2}}
\newcommandx{\jf}[2][1=]{\todo[inline,linecolor=blue,backgroundcolor=black!15,bordercolor=black,#1]{\textbf{JF:} #2}}

\makeatletter
\renewcommand\subsubsection{\@startsection{subsubsection}{3}{\z@}                       {-6\p@ \@plus -1\p@ \@minus -1\p@}                       {-0.5em \@plus -0.22em \@minus -0.1em}                       {\normalfont\normalsize\bfseries\boldmath}}
\makeatother
 
\begin{document}

\mainmatter  
\title{EMME: a formal tool for \\ \textbf{E}CMAScript
  \textbf{M}emory \textbf{M}odel \textbf{E}valuation}

\titlerunning{\textbf{E}CMAScript \textbf{M}emory \textbf{M}odel
  \textbf{E}valuator}

\author{Cristian Mattarei\thanks{This work was supported by a research grant from Google.}
  \and Clark Barrett
  \and Shu-yu Guo
  \and \\Bradley Nelson
  \and Ben Smith
  \and JF Bastien
}
\authorrunning{Mattarei, Barrett, et al.}

\institute{Stanford University, Stanford, CA, USA\\
\mailsa\\
Mozilla, Mountain View, CA, USA\footnote{At the time this work was done}\\
\mailsc\\
Google Inc., Mountain View, CA, USA\\
\mailsd\\
Apple Inc., Cupertino, CA, USA\\
\mailse
}

\toctitle{Lecture Notes in Computer Science}
\tocauthor{Authors' Instructions}
\maketitle
\vspace{-10pt}

\begin{abstract}
Nearly all web-based interfaces are written in JavaScript. Given its
prevalence, the support for high performance JavaScript code is
crucial. The ECMA Technical Committee 39 (TC39) has recently extended
the ECMAScript language (i.e., JavaScript) to support shared memory
accesses between different threads. The extension is given in terms of
a natural language memory model specification.
In this paper we describe a formal approach for validating both the
memory model and its implementations in various JavaScript engines. We
first introduce a formal version of the memory model and report
results on checking the model for consistency and other properties. We
then introduce our tool, EMME, built on top of the Alloy analyzer,
which leverages the model to generate all possible valid executions of
a given JavaScript program. Finally, we report results using EMME
together with small test programs to analyze industrial JavaScript
engines. We show that EMME can find bugs as well as missed
opportunities for optimization.
\end{abstract}

\section{Introduction}\label{sec:introduction}

As web-based applications written in JavaScript continue to increase in
complexity, there is a corresponding need for these applications to interact
efficiently with modern hardware
architectures. Over the last decade, processor architectures have
moved from single-core to multi-core, with the latter
now present in the vast majority of both desktop and
mobile platforms. In 2012, an extension to JavaScript
was standardized\cite{WWK12}
which supports the creation of multi-threaded parallel
Web Workers with message-passing.  More recently, the committee responsible for
JavaScript standardization extended the language to support shared
memory access~\cite{ESM16}. This extension integrates a new datatype called
\emph{SharedArrayBuffer} which allows for concurrent memory accesses,
thus enabling more efficient multi-threaded program interaction.

Given a multi-threaded program that uses shared memory, there can be several
possible valid executions of the program, given that reads
and writes may concurrently operate on the same shared
memory and that every thread can have a different view of
it. However, not all behaviors are allowed, and the separation between
valid and invalid behaviors is defined by a \emph{memory model}. In one common approach,
memory models are specified using axioms, and the
correctness of a program execution is determined by checking its
consistency with the axioms in the memory model.
Given a set of memory operations (i.e., reads and
writes) over shared memory, the memory model defines which
combinations of written values each read event can
observe. Because many different programs can have the same
behaviors, the memory model is also particularly important for helping to
determine the set of possible optimizations that a compiler can apply to
a given program.
As an example, a memory model could specify that the only allowed multi-threaded
executions are those that are equivalent to a sequential program composed of some
interleaving of the events in each thread. This model is the most stringent
one and is called sequential consistency.  With this approach,
all threads observe the same total order of events.  However, this model
has significant performance limitations. In particular, it
requires all cores/processors to synchronize their local cache with each other
in order to maintain a coherent order of the memory events.
In order to overcome such limitations, weaker memory models have been
introduced.  The ECMAScript \memorymodel is a weak model.

Memory models are notoriously challenging to analyze with conventional
testing alone, due to their non-intuitive semantics and formal axiomatic definitions.
As a result, formal methods are frequently used in order to verify and validate
the correctness of memory models~\cite{Jade10,Atig10,Batty11,Ten14,Batty15}.
Some of these models apply to instruction set architectures, whereas others
apply to high-level programming languages.
In this work, we use formal methods to validate the ECMAScript \memorymodel and
to analyze the correctness and performance of different implementations of
ECMAScript engines. JavaScript is usually regarded as a high-level programming
language, but its memory model is decidedly low-level and more closely matches
that of instruction set architectures than that of other languages.
The analyses that we provide are based on a formalization of the memory model
using the Alloy language~\cite{alloy}, which is then combined with a formal
translation of the program to be analyzed in order to compute its set of valid
executions.
This result can then be used to automatically generate litmus tests that can be run
on a concrete ECMAScript engine, allowing the developers to
evaluate its correctness. The concrete executions observed when running
the ECMAScript engine can either be a subset of, be equivalent to, or be a superset of
the valid executions. Standard litmus test analyses usually
target the latter case (incorrect engine behavior),
providing little information in the other cases.
However, when the concrete engine's observed executions are a relatively small subset of the valid executions,
(e.g., 1/5 the size), this can indicate a missed opportunity for code
optimization. As part of our work, we introduce a novel approach in such cases
that is able to identify specific predicates over the memory model that are
always consistent with the executions of the concrete engine, thus
providing guidance about where potential optimization opportunities might exist.

The analyses proposed in this paper have been implemented in a tool
called \textbf{E}CMAScript \textbf{M}emory \textbf{M}odel
\textbf{E}valuator (EMME), which has been used to validate the
memory model and to test the compliance of all
major ECMAScript engines, including Google's V8~\cite{V8}, Apple's
JSC~\cite{JSC}, and Mozilla's SpiderMonkey~\cite{SpiderMonkey}.

The rest of the paper is organized as follows:
\begin{inparaenum}[]
\item Section~\ref{sec:related_works} covers
  related work on formal analysis of memory models;
\item Section~\ref{sec:ecma_mm} describes the \ES \memorymodel and its
  formal representation;
\item Section~\ref{sec:formal_analyses} characterizes the analyses
  that are presented in this paper;
\item Section~\ref{sec:sat_approach} provides an overview of the Alloy
  translation;
\item Section~\ref{sec:implementation} concentrates on the tool
  implementation and the design choices that were made;
\item Section~\ref{sec:evaluation} provides an evaluation of the performance of the
  different techniques proposed in this paper;
\item Section~\ref{sec:results_analysis} describes the results of the
  analyses performed on the ECMAScript \memorymodel and several specific
  engine implementations; and
\item Section~\ref{sec:conclusion} provides concluding remarks.
\end{inparaenum}

\section{Related Work}
\label{sec:related_works}

Most modern multiprocessor systems implement relaxed memory
models, enabling them to deliver better performance when compared to more
strict models.  Well known approaches such as Sequential
Consistency (SC), Processor Consistency (PC), Relaxed-Memory Order
(RMO), Total Store Order (TSO), and Partial Store Order (PSO) are
mainly directed towards relaxing the constraints on when read and
write operations can be reordered.

The formal analysis of weak memory model hardware implementations
has typically been done using SAT-based
techniques~\cite{Atig10,Burckhardt07}. In \cite{Jade10}, a formal
analysis based on Coq is used in order to evaluate SC, TSO, PSO, and
RMO memory models. The DIY tool developed in \cite{Jade10} generates
assembly programs to run against Power and x86 architectures. In contrast,
in this work we concentrate on the analysis of the ECMAScript memory
model, assuming the processor behavior is correct.

MemSAT~\cite{Torlak10} is a formal tool, based on Alloy~\cite{alloy},
that allows for the verification of axiomatic memory models. Given a
program enriched with assertions, MemSAT finds a trace execution (if
it exists) where both assertions and the axioms in the memory model
are satisfied.

An analysis of the C++ memory model is presented in
\cite{Batty15}. The formalization is based on the LEM
language~\cite{Owens11}, and the CPPMem software provides all possible
interpretations of a C/C++ program consistent with the memory
model. More recently, an approach based on Alloy and oriented towards
synthesizing litmus tests is proposed in \cite{nvidia-alloy}.

In this paper, we build on ideas present in MemSAT and CPPMem to
build a tool for \JS.  Our EMME tool can provide the set of valid
executions for a given input \JS program, and it can also
generate litmus tests suitable for evaluating the correctness of
\JS engine implementations. In contrast to previous work, we also
analyze situations where the litmus tests
provide correct results but expose a discrepancy between
the number of observed behaviors in the implementation and what is possible
given the specification.

\section{The ECMAScript Memory Model}
\label{sec:ecma_mm}

\begin{figure}[t]
\centering
\begin{minipage}{.59\textwidth}
\[
\underbrace{\overbrace{{\footnotesize\texttt{init x = 0}}}^\text{$ev_1W^1$}}_\text{Thread 1}{\footnotesize\texttt{ | }}
\underbrace{\overbrace{{\footnotesize\texttt{x-I8[0] = 1}}}^\text{$ev_2W^2$}{\footnotesize\texttt{ ; print(}}
  \overbrace{{\footnotesize\texttt{x-I16[0]}}}^\text{$ev_3R^2$}\footnotesize\texttt{)}}_\text{Thread 2}{\footnotesize\texttt{ | }}
\]
\vspace{-05pt}
\[
\underbrace{{\footnotesize\texttt{ite(}}
\overbrace{{\footnotesize\texttt{x-I8[0]}}}^\text{$ev_4R^3$}
{\footnotesize\texttt{==~1, }}
\overbrace{{\footnotesize\texttt{x-I8[0] = 3}}}^\text{$ev_5W^3$}{\footnotesize\texttt{, }
\overbrace{{\footnotesize\texttt{x-I8[1] = 3}}}^\text{$ev_6W^3$}}{\footnotesize\texttt{)}}}_\text{Thread 3}
\]
\caption{Concurrent Program Example}
\label{fig:simple_program_e1}
\end{minipage}\begin{minipage}{.41\textwidth}
  \centering
  \vspace{16pt}
  \includegraphics[width=.7\textwidth]{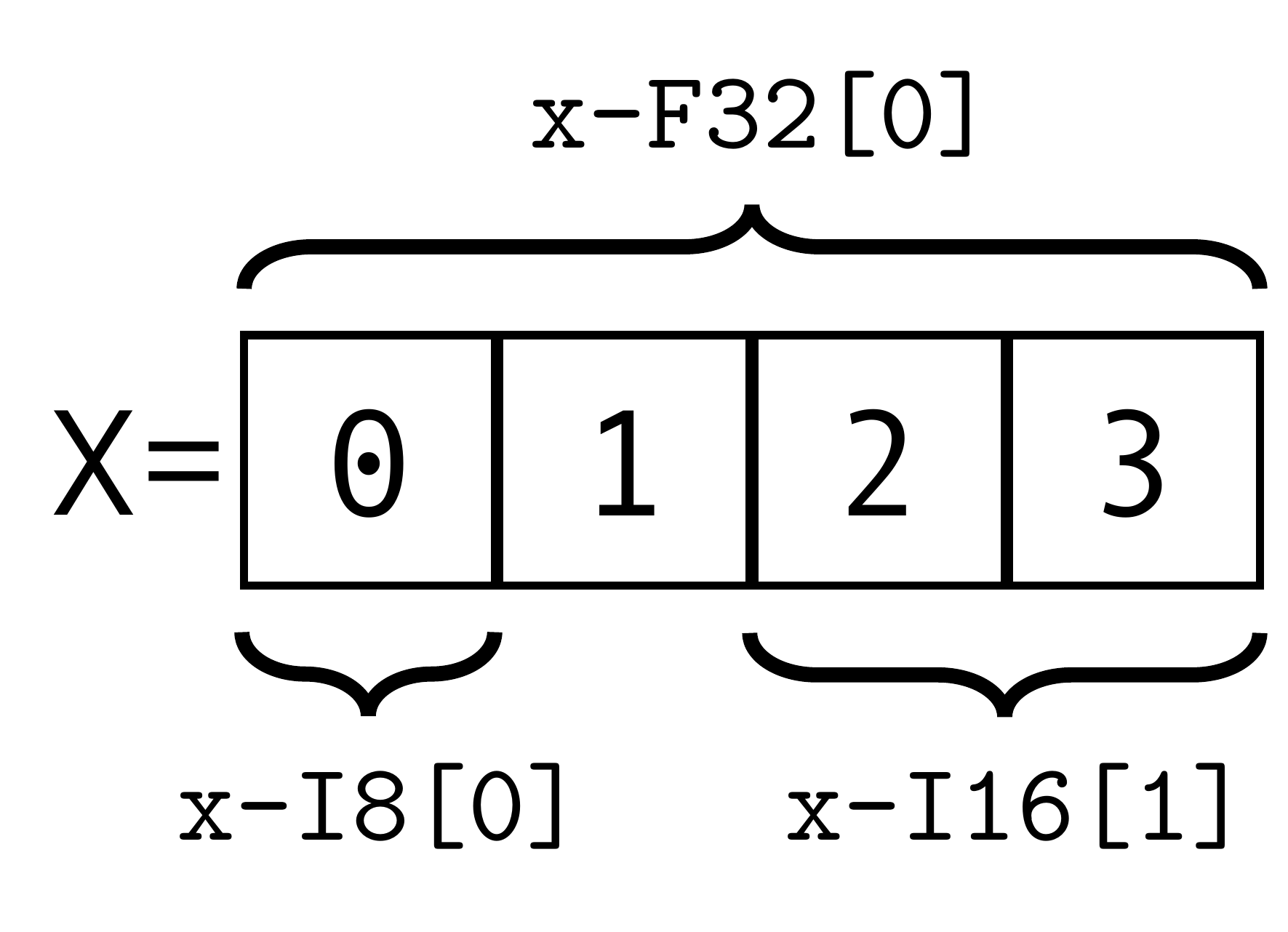}
  \vspace{02pt}
  \caption{Shared Memory Views}
  \label{fig:memory_views}
\end{minipage}
\end{figure}

The objective of the ECMAScript \memorymodel is to precisely define
when an execution of a concurrent program that relies on shared memory
is valid. From the point of view of the \memorymodel, a JavaScript program can
be abstracted as a set of threads, each of them composed of an ordered set of shared
memory events. Each memory event has a set of attributes that specify
its: operation (\emph{Read}, \emph{Write}, or \emph{ReadModifyWrite});
ordering (\emph{SeqCst}, \emph{Unordered}, or \emph{Init}); tear type
(whether a single read operation can read from two different writes to the same location);
(source or destination) memory block and address; payload value; and
modify operation (in the case of a \emph{ReadModifyWrite}).  The
shared memory is essentially an array of bytes, and a memory operation
reads, writes, or modifies it.  In these operations, the bytes can be
interpreted either as \emph{signed/unsigned integer} values or as
\emph{floating point} values. For
instance, in Figure~\ref{fig:memory_views},
the notation \texttt{x-I16[1]} represents an access to the
memory block \texttt{x} starting at index 1, where the bytes are interpreted as 16-bit signed integers
(i.e., \texttt{I16}), while \texttt{x-F32[0]} stands for a 32-bit
floating point value starting at byte 0.

Formally, a program is defined as a set of events $E$ and a partial
order between them, namely the \emph{Agent Order}, that encodes the
thread structure. For the example in
Figure~\ref{fig:simple_program_e1}, the set of events is defined as $E
= \{ev_1W^1$, $ev_2W^2$, $ev_3R^2$, $ev_4R^3$, $ev_5W^3$, $ev_6W^3\}$,
with agent order $\text{AO} = AO^1 \cup AO^2 \cup AO^3$, where $AO^1$,
$AO^2$, and  $AO^3$ are the agent orders for each thread:
$\text{AO}^1 = \{\}$, $\text{AO}^2 = \{(ev_2W^2, ev_3R^2)\}$, and
$\text{AO}^3 = \{(ev_4R^3, ev_5W^3)$, $(ev_4R^3, ev_6W^3)$, $(ev_5W^3,
ev_6W^3)\}$.

The execution semantics of a program is given by the
\RBF (RBF) relation, a trinary relation which relates two events and
a single byte index $i$, with the interpretation that the first event reads the byte at index
$i$ which was written by the second event.  Looking again at the example in
Figure~\ref{fig:simple_program_e1}, one of the possible valid
assignments to the RBF relation is $\{(ev_4R^3, ev_1W^1, 0)$,
$(ev_3R^2, ev_2W^2, 0)$, $(ev_3R^2, ev_6W^3, 1)\}$, meaning that the
\emph{Read} event $ev_4R^3$ reads byte 0 from $ev_1W^1$ (taking the
else branch), and $ev_3R^2$ reads byte 0 from $ev_2W^2$ and 1 from
$ev_6W^3$.

The combination of a (finite) set of events $E = \{e_1, \ldots,
e_n\}$, an agent order $AO \in E \times E$, and a \RBF $RBF \in E
\times E \times \mathbb{N}$ relation identify a \emph{Candidate
  Execution}, and the purpose of the \memorymodel is to partition this set
into \emph{Valid} and \emph{Invalid} executions. The separation is
defined as a formula that is satisfiable if
and only if the \emph{Candidate Execution} is \emph{Valid}. Given a
\emph{Candidate Execution}, the \memorymodel constructs a set of
supporting relations in order to assess its validity:
\begin{itemize}
\item \emph{Reads From} (RF): a binary relation that generalizes RBF
  by dropping the byte location;
\item \emph{Synchronizes With} (SW): the synchronization relation
  between sequentially consistent writes and reads;
\item \emph{Happens Before} (HB): a partial order relation between all
  events;
\item \emph{Memory Order} (MO): a total order relation between
  sequentially consistent events.
\end{itemize}

\noindent
Finally, a \emph{Candidate Execution} is valid when the following
predicates hold:
\begin{itemize}
\item \emph{Coherent Reads} (CR): RF and HB relations are consistent;
\item \emph{Tear Free Reads} (TFR): for reads and writes for which the tear
  attribute is false, a single read event cannot read from two different write events
  (both of which are to the same memory address);
\item \emph{Sequential Consistent Atomics} (SCA): the MO relation is
  not empty.
\end{itemize}

\subsection{Formal Representation}
\label{sec:ecma_mm_fr}

The formalization of the \ES \memorymodel is based on the formal
definition of a {\em Memory Operation}, shown in 
Definition~\ref{def:memory_operation}.

\begin{definition}[Memory Operation]
  \label{def:memory_operation}
  A Memory Operation is a tuple $\langle ID$, $O$, $T$, $R$, $B$, $M$,
  $A \rangle$ where:
  \begin{itemize}
  \item $ID$ is a unique event identifier;
  \item $O \in \{\text{Read (R)}, \text{Write (W)}, \text{ReadModifyWrite (M)}\}$ is the operation;
  \item $T \in \mathbb{B}$ is the Tear attribute;
  \item $R \in \{\text{Init (I)}, \text{SeqCst (SC)}, \text{Unordered
    (U)}\}$ is the order attribute;
  \item $B$ is the name of a Shared Data Block;
  \item $M$ is a set of integers
    representing the memory addresses in $B$ accessed by the operation $O$, with the
    requirement that $M = \{i \in \mathbb{N}\ |\ ByteIndex \leq i <
ByteIndex + ElementSize \}$, for some
$ByteIndex, ElementSize \in \mathbb{N}$
  \item $A \in \mathbb{B}$ is an Activation attribute.
  \end{itemize}
\end{definition}

\noindent
Note that this definition differs slightly from the one used
in \cite{ESM16} (though the underlying semantics are the same).  The
differences make the model easier to reason about formally and include:
\begin{itemize}
  \item In \cite{ESM16}, the memory address range for an operation is
    represented by two numbers, the \emph{ByteIndex} and the \emph{ElementSize}, whereas in
    Definition~\ref{def:memory_operation}, we represent the memory address
    range explicitly as a set of bytes (which must contain some set of
    consecutive numbers, so the two representations are equivalent).
    This representation allows for a simpler encoding of some operators like
    computing the intersection of two address ranges.
  \item Definition~\ref{def:memory_operation} omits the payload
and modify operation attributes, as these are only needed to compute the
concrete value(s) of the data being read or written.  The formal model does not
need to reason about such concrete values in order to partition candidate
executions into valid and invalid ones.  Furthermore, for any specific
candidate execution of a JavaScript program, these values can be computed from
the original program using the RBF relation.
\item The activation attribute $A$ is an extension used to encode whether an
  event should be considered active based on the control flow path taken in
  an execution.  In particular, we model \emph{if-then-else} statements by
  enabling or disabling the events in the then and else branches depending on
  the value of the condition.
\end{itemize}

All relations in \cite{ESM16} (i.e., RBF, RF, SW, HB, and MO) are
included in the formal model, and their semantics are defined using set operations,
while the predicates (i.e., CR, TFR, and SCA)
are expressed as formulas. The resulting
formulation of the \memorymodel, combining all constraints and
predicates, is shown in Equation (\ref{eq:mm}).  Details of our implementation
of this formulation are given in Section~\ref{sec:sat_approach}.
{\small
  \begin{align}
    MM&(E, AO, RF, RBF, SW, HB, MO) := \varphi_{RBF}(RBF, E) \wedge \varphi_{RF}(RF, E, RBF) \wedge \nonumber \\
    &\varphi_{SW}(SW, E, RF) \wedge \varphi_{HB}(HB, E, AO, SW) \wedge \varphi_{MO}(MO, E, HB, SW) \wedge \nonumber \\
    &CR(E, HB, RBF) \wedge TFR(E, RF) \wedge SCA(MO)
    \label{eq:mm}
  \end{align}
}
 
\section{Formal Analyses}
\label{sec:formal_analyses}

The design and development of a critical (software or hardware) system often follows a
process in which high-level requirements (such as the standards committee's
specification of the memory model) are used to guide an actual implementation. This process can be
integrated with different formal analyses to ensure that the result is
a faithful implementation with respect to the requirements.  In this section, we describe the
set of analyses that we used to validate the
requirements and implementations of the ECMAScript
\memorymodel.  Results of our analyses are reported in Section~\ref{sec:results_analysis}.

\subsection{Formal Requirements Validation}

The \ES \memorymodel defines a set of \emph{constraints} which together make up
a formula (Equation (\ref{eq:mm})).  The solutions of this formula are the valid executions.
The \memorymodel also lists a number of \emph{assertions}, formulas that are
expected to be true in every valid execution (and thus must follow from the constraints).
Complete formal requirements validation would require checking
two things: (i) the constraints are consistent with each other, i.e. they contain
no contradictions; and (ii) each assertion is logically entailed by the set of
constraints in the \memorymodel.  However, because we used Alloy (see
Section~\ref{sec:sat_approach}) we were unable to show full logical
entailment, as Alloy can only reason about a finite number of events.  So we
instead showed that for finite sets of events up to a certain size, (i) and
(ii) hold.  In future work, we plan to explore using an SMT solver to see if we
can prove unbounded entailment in some cases.  When (i) or (ii) do not hold,
there is a bug in either the requirements or the formal modeling of the
requirements.  To help debug problems with (i), we used the unsat core feature of Alloy,
which identifies a subset of the constraints that are inconsistent.  To further
aid debugging, we labeled each constraint $c_i$ with a
Boolean activation variable $av_i$ (i.e. we replaced $c_i$ with $(av_i \to
c_i) \wedge av_i$).  This allowed us to inspect the unsat core for activation
variables and immediately discern which constraints were active in producing
the unsatisfiable result.

\subsection{Implementation Testing}

The \emph{Implementation testing} phase analyzes whether a
specific JavaScript engine correctly implements the \ES
\memorymodel. In particular, given a program with shared memory
operations, we generate:
\begin{inparaenum}[1)]
\item the set of valid executions,
\item a litmus test, and
\item behavioral coverage constraints.
\end{inparaenum}

\subsubsection{Valid Executions}
This analysis lists all of (and only) the behaviors that the (provided)
program can exhibit that are consistent with the \memorymodel specification. The
encoding of the problem is based on the following definition:
{\small
  \begin{align*}
    \text{VE(E, AO)} := \{&(\text{RBF, HB, MO, SW})\ |\ \\
    &\text{MM(E, AO, RF, RBF, SW, HB, MO)}\ \text{is SAT}\}
\end{align*}}
where VE(E, AO) is the complete (and finite because the program itself is finite) set of possible assignments
to the RBF, HB, MO, and SW relations.  Each assignment corresponds to a
\emph{valid} execution.

\subsubsection{Litmus Tests}
\emph{Litmus test generation} uses the generated list of valid
executions to construct a JavaScript program
enriched with an assertion that is violated if the output of the
program does not match any of the valid executions. A litmus
test is executed multiple times (e.g., millions), in order to increase
the chance of exposing a problem if there is one.

The result of running a litmus test many times can (in general) have one of
three outcomes: the assertion is violated at least once, the assertion is not
violated and all possible executions are observed, and the assertion is not
violated and only some of the possible executions are observed. More specifically, given a program $P$, the set of
its valid executions $\text{\emph{VE}}(P)$, and the set of concrete
executions $E_{N}(P)$ (obtained by running the JavaScript program on engine $E$ some
number of times $N$), the possible results can be
respectively expressed as $E_{N}(P) \setminus \text{\emph{VE}}(P)
\neq \emptyset$, $E_{N}(P) = \text{\emph{VE}}(P)$, and $E_{N}(P)
\subset \text{\emph{VE}}(P)$.

\subsubsection{Behavioral Coverage Constraints}

Though they can expose bugs, the litmus tests do not provide a guarantee of
implementation correctness. In fact, even when a ``bug'' is found,
it could be that the specification is too tight (i.e., it is incompatible with some
intended behaviors) rather than that the implementation wrong.
On the other hand, when $E_{N}(P) \subset \text{\emph{VE}}(P)$,
and especially if the cardinality of $E_{N}(P)$ is significantly smaller than that
of $\text{\emph{VE}}(P)$, it might be the case that the
implementation is too simple: it is not taking sufficient advantage of the weak memory model and
is therefore unnecessarily inefficient.

Whenever $E_{N}(P) \subset \text{\emph{VE}}(P)$, this situation can be
analyzed by the generation of \emph{Behavioral Coverage Constraints}.
The goal of this analysis is to synthesize the formulae $\Sigma_{OBS}$
and $\Sigma_{UNOBS}$, for observed and unobserved outputs, that
restrict the behavior of the memory model in order to match $E_{N}(P)$
and $\text{\emph{VE}}(P)\setminus E_{N}(P)$.

Our approach to doing this relies on first choosing a set $\Pi = \{\pi_1, \dots, \pi_n\}$ of predicates over
which the formula will be constructed.  One choice for $\Pi$ might be all
atomic predicates appearing in Equation (\ref{eq:mm}).  Now, let $\Delta(\Pi)$
be the set of all cubes of size $n$ over $\Pi$.  Formally,
\[\Delta(\Pi) = \{l_1\wedge \dots \wedge l_n\ |\ \forall\,1\le i\le n.\: l_i\in\{\pi_i,\neg\pi_i\}\}.\]
Further, define the observed and unobserved executions as:
\[\begin{array}{lcl}
EX_{OBS} &= &\bigvee_{\langle\text{RBF, HB, MO, SW}\rangle\in E_{N}(P)}(RBF\wedge HB \wedge MO \wedge SW)\\
EX_{UNOBS} &= &\bigvee_{\langle\text{RBF, HB, MO, SW}\rangle\in VE(P)\setminus E_{N}(P)}(RBF\wedge HB \wedge MO \wedge SW)
\end{array}\]
We compute those cubes in $\Delta(\Pi)$ that are
consistent with the observed and unobserved executions as follows:
\[\begin{array}{lcl}
\delta_{OBS}(\Pi) &= &\{\delta \in \Delta(\Pi)\ |\ MM \wedge EX_{OBS} \wedge \delta \text{ is satisfiable}\}\\
\delta_{UNOBS}(\Pi) &= &\{\delta \in \Delta(\Pi)\ |\ MM \wedge EX_{UNOBS} \wedge \delta \text{ is satisfiable}\}
\end{array}\]
The cubes are then combined to generate the formulae for matched and unmatched executions:
\[\Sigma_{OBS} = \bigvee_{\delta\in\delta_{OBS}} \delta, ~~~~\Sigma_{UNOBS} = \bigvee_{\delta\in\delta_{UNOBS}}\delta.\]

For example, let $(\text{R2H} := \forall_{e_1,e_2 \in E} : RF(e_1,e_2)
\to HB(e_1,e_2)) \in \Pi$ be a predicate expressing that every tuple
in \RF is also in \HB. If the behavioral coverage constraints analysis
generates $\Sigma_{OBS} = \text{R2H}$ and $\Sigma_{UNOBS} =
\neg \text{R2H}$, it means that the \JS engine always aligns the read from
relation with the HB relation, thus identifying a possible path for
optimization in order to take advantage of the (weak) memory model.

\section{Alloy Formalization}
\label{sec:sat_approach}

Alloy is a widely used modeling language that can be used to describe
data structures. The Alloy language is based on relational algebra
and has been successfully used in many applications, including the analysis of
memory models~\cite{nvidia-alloy}.

We used Alloy to formalize the memory model discussed in Section~\ref{sec:ecma_mm_fr}.
We followed the formalization given in Definition~\ref{def:memory_operation},
using sets and relations to represent each concept.\footnote{\scriptsize{The
complete Alloy model is available at \url{https://github.com/FMJS/EMME/blob/master/model/memory\_model.als}}}. For
instance, an \texttt{operation\_type} is defined as an (abstract) set
with three disjoint subsets (\texttt{R} for \emph{Read},
\texttt{W} for \emph{Write}, and \texttt{M} for
\emph{ReadModifyWrite}), one for each possible operation.  In contrast,
\texttt{blocks} and \texttt{bytes} are represented as sets. A memory operation
is modeled as a relation which links all of the attributes necessary
to describe a memory event.

\begin{figure}[t]
  \centering
  \begin{minipage}{1.0\textwidth}
    {\scriptsize
      \noindent\rule{1.0\textwidth}{0.4pt}
      6.3.1.14 happens-before
      \begin{enumerate}[4.]
      \item
        For each pair of events E and D in EventSet(execution):
        \begin{enumerate}[a.]
        \item
          If E is agent-order before D then E happens-before D.
        \item
          If E synchronizes-with D then E happens-before D.
        \item ...
        \end{enumerate}
    \end{enumerate}}
    \vspace{-12pt}
    \noindent\rule{1.0\textwidth}{0.4pt}
  \end{minipage}
  \caption{Excerpt of the \HB definition~\cite{ESM16}}
  \label{fig:hb_excerpt}
  \vspace{-12pt}
\end{figure}

The formalization of a natural language specification usually
requires multiple attempts and iterations before the intended
semantics become clear. In the case of the \ES \memorymodel, this process
was crucial for disambiguating some of the stated constraints. An example
is the \HB relation. Figure~\ref{fig:hb_excerpt} shows
an excerpt of its definition, expressing how it is related to the \AO and
\SW relations.  One might expect that the formal interpretation would be
something like:
$\forall\,(e_1,e_2).\:(\text{AO}(e_1,e_2) \rightarrow \text{HB}(e_1,e_2)) \wedge
(\text{SW}(e_1,e_2) \rightarrow \text{HB}(e_1,e_2)) \wedge (\ldots)$

\hspace{-18pt}
\begin{minipage}{\linewidth}
\begin{lstlisting}[caption=Excerpt of the \HB definition, label=all:hb_excerpt]
  fact hb_def {all ee,ed : mem_events | Active2 [ee,ed] =>
    (HB [ee,ed] <=> ((ee != ed) and (AO [ee,ed] or SW [ee,ed] or ... )))}
\end{lstlisting}
\end{minipage}

However, further analysis and discussions with the people responsible
for the \memorymodel revealed that the correct interpretation is:
$\forall\,(e_1,e_2).\:\text{HB}(e_1,e_2) \leftrightarrow
(\text{AO}(e_1,e_2) \vee \text{SW}(e_1,e_2) \vee \ldots)$.  The
Alloy formalization of the \HB relation is shown in
Figure~\ref{all:hb_excerpt}.  The \texttt{Active2} predicate evaluates to true
when both events are active.

Once the \memorymodel has been formalized, the next step is to
combine it with the encoding of the program under analysis. This requires
modeling the memory events present in each thread. In the Alloy model, each event in a program
extends the set of memory events, and its values
are defined as a series of facts.  Figure~\ref{all:id5simple03} shows an
example of the Alloy model for the event $ev_5W^3$ from Figure~\ref{fig:simple_program_e1}.
A notable aspect of
this example is the fact that its activation is
dependent on the value of \texttt{id1\_cond} which symbolically
represents the condition of the \ITE statement.

\hspace{-18pt}
\begin{minipage}{\linewidth}
\begin{lstlisting}[basicstyle=\tiny, caption=event $ev_5W^3$ encoding (w.r.t. Figure~\ref{fig:simple_program_e1}), label=all:id5simple03]
one sig ev5_W_t3 extends mem_events{}
fact ev5_W_t3_def {(ev5_W_t3.O = W) and
                   (ev5_W_t3.T = NT) and
                   (ev5_W_t3.R = U) and
                   (ev5_W_t3.M = {byte_0}) and
                   ((ev5_W_t3.A = ENABLED) <=> ((id1_cond.value = TRUE))) and
                   (ev5_W_t3.B = x)}
fact ev5_W_t3_in_mem_events {ev5_W_t3 in mem_events}
\end{lstlisting}
\end{minipage}

\section{Implementation}
\label{sec:implementation}

The techniques descriped in this paper have been implemented in a
tool called EMME: \textbf{E}CMAScript \textbf{M}emory \textbf{M}odel
\textbf{E}valuator~\cite{EMME}. The tool is written in Python, is open
source, and its usage is regulated by a modified BSD license. The
input to EMME is a program with shared memory accesses.
The tool interacts with the Alloy Analyzer~\cite{alloy_analyzer} to perform the
formal analyses described in
Section~\ref{sec:formal_analyses}, which include the enumeration of valid executions and
the generation of behavioral coverage constraints.

\subsubsection{Input Format and Encoding}

\begin{wrapfigure}{r}{0.5\textwidth}
  \vspace{-28pt}
  \centering
  \begin{minipage}{0.4\textwidth}
\begin{lstlisting}[basicstyle=\tiny]
var x = new SharedArrayBuffer();

Thread t1 {
  x-I8[0] = 1;
  print(x-I16[0]);
}

Thread t2 {
  if (x-I8[0] == 1) {
     x-I8[0] = 3;
  } else {
     x-I8[1] = 3;
  }
}
\end{lstlisting}
  \end{minipage}
  \caption{EMME input for the program from Figure~\ref{fig:simple_program_e1}.}
  \label{fig:simple_program_e1_be}
  \vspace{-20pt}
\end{wrapfigure}

The input format of EMME uses a simplified JavaScript-like syntax. It supports the
definition of \emph{Read}, \emph{Write}, and \emph{ReadModifyWrite}
events, allows events to be atomic or not atomic, and supports operations on
integer or floating point values. The input format also supports
\ITE and bounded \FOR statements, as well as parametric values. An example of
an input program is shown in Figure~\ref{fig:simple_program_e1_be}.
The program is encoded in Alloy and combined with the memory
model in order to provide the input formula for the formal analyses.

\subsubsection{Generation of All Valid Executions}
\label{sec:all_valid}

The generation of all valid executions is computed by using Alloy to solve the
AllSAT problem.
In this case, the distinguishing models of the formula are
the assignments to the RBF relation. Thus, after each
satisfiability check iteration of the Alloy Analyzer, an additional
constraint is added in order to block the current assignment to the RBF
relation. This procedure is performed until the model becomes
unsatisfiable.

As described in Section~\ref{sec:ecma_mm_fr}, our formal model does not
encode the concrete values of each memory operation; thus, the
extraction of a valid execution, given a satisfiable assignment to the
formula, requires an additional step. This step is to
to reconstruct the values of each read or modify operation
based on the program and the assignment to the RBF relation. For example, given
the program in Figure~\ref{fig:simple_program_e1}, and assuming that
the RBF relation contains the tuples $(ev_3R^2, ev_2W^2, 0)$ and
$(ev_3R^2, ev_6W^3, 1)$, the reconstruction of the value read by
$ev_3R^2$ depends on the fact that $ev_2W^2$ writes 1 with an 8-bit
integer encoding at position 0, while $ev_6W^3$ writes 3 at position
1. The composition of byte 0 and byte 1 from those two writes is the
input for the decoding of a 16-bit integer for the event $ev_3R^2$,
resulting in a read of the value 769. Clearly, each event could also have
a different size and format (i.e., integer, unsigned integer, or float);
thus, the reconstruction of the correct value must also take this
into account.

When interpreting a program containing \ITE statements,
the possible outcomes must be filtered to exclude executions that break the
semantics of \ITE.  In particular, it
might be the case that the Boolean condition in the model does not
match the concrete value, given the read values. For instance,
consider the example in Figure~\ref{fig:simple_program_e1_be} in which
the conditional is encoded as a Boolean variable \texttt{id1\_cond}
representing the statement \texttt{x-I8[0] == 1}. However, the tool may
assign \texttt{id1\_cond} to false even though the event
\texttt{x-I8[0]} turns out to read a value different from 1 based on
the information in the RBF relation. In this case, this execution
is discarded since it is not possible given the semantics of the
\ITE statement.

\paragraph{Graph Representation of the Results}

\begin{figure}[t]
     \begin{center}
        \subfigure[Interpretation 1 (\emph{THEN})]{          \label{fig:gmm1}
          \hspace{20pt}
            \includegraphics[width=0.40\textwidth]{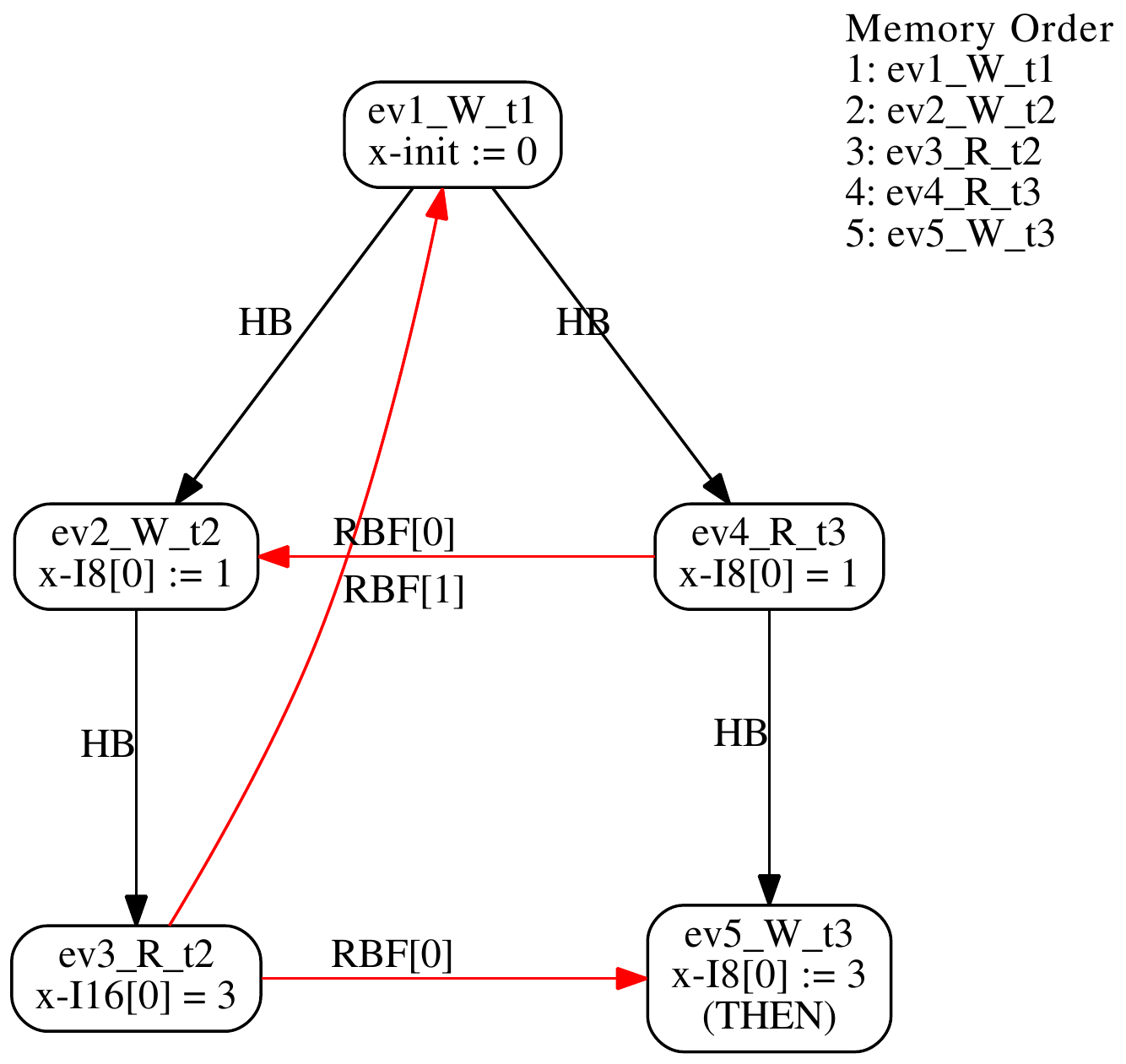}
        }        \subfigure[Interpretation 2 (\emph{ELSE})]{            \label{fig:gmm2}
          \hspace{20pt}
            \includegraphics[width=0.40\textwidth]{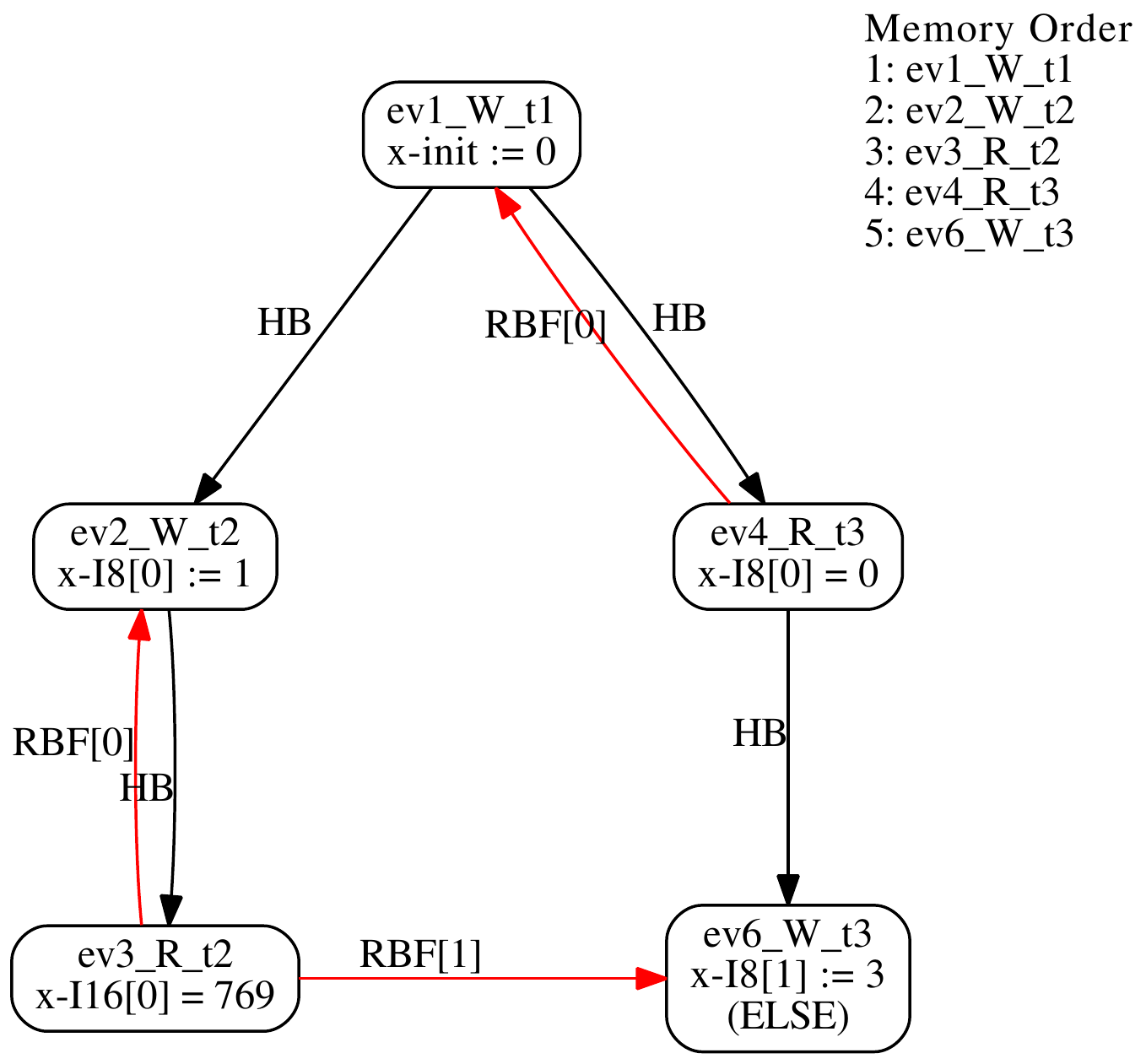}
        }    \end{center}
     \vspace{-07pt}
    \caption{        Memory Model interpretations of the program in Figure~\ref{fig:simple_program_e1_be}.
     }    \label{fig:mminterps}
        \vspace{-08pt}
\end{figure}

For each valid execution, EMME will produce a graphviz file that
provides a graphical representation of the assignments to main
relations and read values. An example of this graphical
representation is shown in Figure~\ref{fig:mminterps}. The default
setup removes some redundant information such as the explicit
transitive closure of the HB relation, while RF and AO are not
represented, and the total order MO is reported in the top right
corner. Black arrows are used to represent the HB relation, while red
and blue are respectively used for RBF and
SW. Figure~\ref{fig:gmm1} represents an execution where event
\texttt{ev4\_R\_t3} reads value 1 from \texttt{ev2\_W\_t2}, thus
executing the \emph{THEN} branch in the \ITE statement. In contrast,
Figure~\ref{fig:gmm2} reports an execution where it reads 0, thus taking
the \emph{ELSE} branch.

\paragraph{Litmus Test Generation}

The generation of all valid executions also constructs a JavaScript
litmus test that can be used to evaluate whether the engine respects the semantics
of the \memorymodel. The structure of the litmus test mirrors that of the input
program, but the syntax follows the official TEST262 ECMAScript conformance
standard~\cite{T262}.

To check whether a test produced a valid result, the results of memory
operations must be collected.  The basic idea consists of printing the values of each read and
collecting them all at the thread level. The main thread is then
responsible for collecting all the results. The sorted report is then
compared with the set of expected outputs using an assertion. Moreover, the
test contains a part that is parsed by the Litmus script, which is
provided along with the EMME tool, and provides a list of expected
outputs. The Litmus script is used to facilitate the execution of
multiple runs of the same test, and it will provide a summary of the
results as well as a warning whenever one of the executions observed
is a not valid according to the standard.

\subsubsection{Generation of the Behavioral Coverage Constraints}

As described in Section~\ref{sec:all_valid}, for each assignment to
the RBF relation, it is possible to construct a concrete value for each
memory event. Thus, for each RBF assignment in a set of valid executions for a
given program, we can determine the output of the corresponding litmus test. Thus, running the litmus
test many times on a \JS engine, it is possible to determine which
assignments to the RBF relation have been matched.  We denote these $\text{MA\_rbf}_1, \ldots,
\text{MA\_rbf}_n$.  The unmatched assignments to RBF can also be determined
simply by removing the matched ones from the set of all valid executions.  We
denote the unmatched ones $\text{UN\_rbf}_1,$ $\ldots,$ $\text{UN\_rbf}_m$.

As described in Section~\ref{sec:formal_analyses}, the generation of 
separation constraints that distinguish between matched and unmatched executions
first requires the definition of a set of predicates $\Pi$. The extraction of the separation
constraints is based on an AllSAT call for matched and unmatched
results. The former is shown in (\ref{eq:separation_ma}), and
consists of extracting all assignments to the predicates $\Pi$ such that
the models of the RBF relation are consistent with $\text{MA\_rbf}_i$.
\begin{equation}
  \label{eq:separation_ma}
  {\small
  \begin{split}
    \text{ALLSAT}_{\Pi}[MM(E, AO, RBF, \ldots) \wedge (E = BE_E) \wedge (AO = BE_{AO}) \wedge \\
  (\bigvee_{i = 1, \ldots, k} RBF = \text{MA\_rbf}_i)]
  \end{split}}
\end{equation}

Similarly, the evaluation for the unmatched executions performs an
AllSAT analysis for the formula reported in
(\ref{eq:separation_un}). The results of these two calls to the solver
produce respectively the formula $\Sigma_{OBS}$ and
$\Sigma_{UNOBS}$ as described in
Section~\ref{sec:formal_analyses}.
\begin{equation}
  \label{eq:separation_un}
  {\small
  \begin{split}
    \text{ALLSAT}_{\Pi}[MM(E, AO, RBF, \ldots) \wedge (E = BE_E) \wedge (AO = BE_{AO}) \wedge \\
  (\bigvee_{i = 1, \ldots, k} RBF = \text{UN\_rbf}_i)]
  \end{split}}
\end{equation}

The results from the two AllSAT queries can then be manipulated using a
BDD~\cite{Bryant:1992:SBM:136035.136043} package that produces in most
cases a smaller formula. After this step, the tool provides
a set of formal comparisons that can be done between these two formulas such as
implication, intersection, and disjunction, in order to understand the
relation between $\Sigma_{OBS}$ and $\Sigma_{UNOBS}$.
 
\section{Experimental Evaluations}
\label{sec:evaluation}

In this section, we evaluate the performance of EMME over a set of
programs, each containing up to 8 memory events. The analyses can be
reproduced using the package available at~\cite{tacasae}.

\paragraph{Programs Under Analysis}

In this work, we rely on programs from previous work \cite{Batty15} as
well as handcrafted and automatically generated programs. The
handcrafted examples are part of the EMME~\cite{EMME} distribution,
and they cover a variety of different configurations with 1 to 8
memory events, if-statements, for-loops, and parametric
definitions. 

The programs from previous work as well as the handcrafted examples
cover an interesting set of examples, but provide no
particular guarantees on the space of programs that are covered. To
overcome this limitation, we implemented a tool that enumerates all possible
programs of a fixed size, thus giving us the possibility of
generating programs to entirely cover the space of configurations,
given a fixed set of events.

The sizes of the programs considered in this evaluation allow us to
cover a representative variety of possible event interactions, while
preserving a reasonable level of readability of the results. In fact,
a program with 8 memory events can have hundreds of valid executions
that often require extensive manual effort to understand.

\paragraph{All Valid Executions}

\begin{wrapfigure}{r}{0.5\textwidth}
  \vspace{-33pt}
     \begin{center}
           \includegraphics[width=0.49\textwidth]{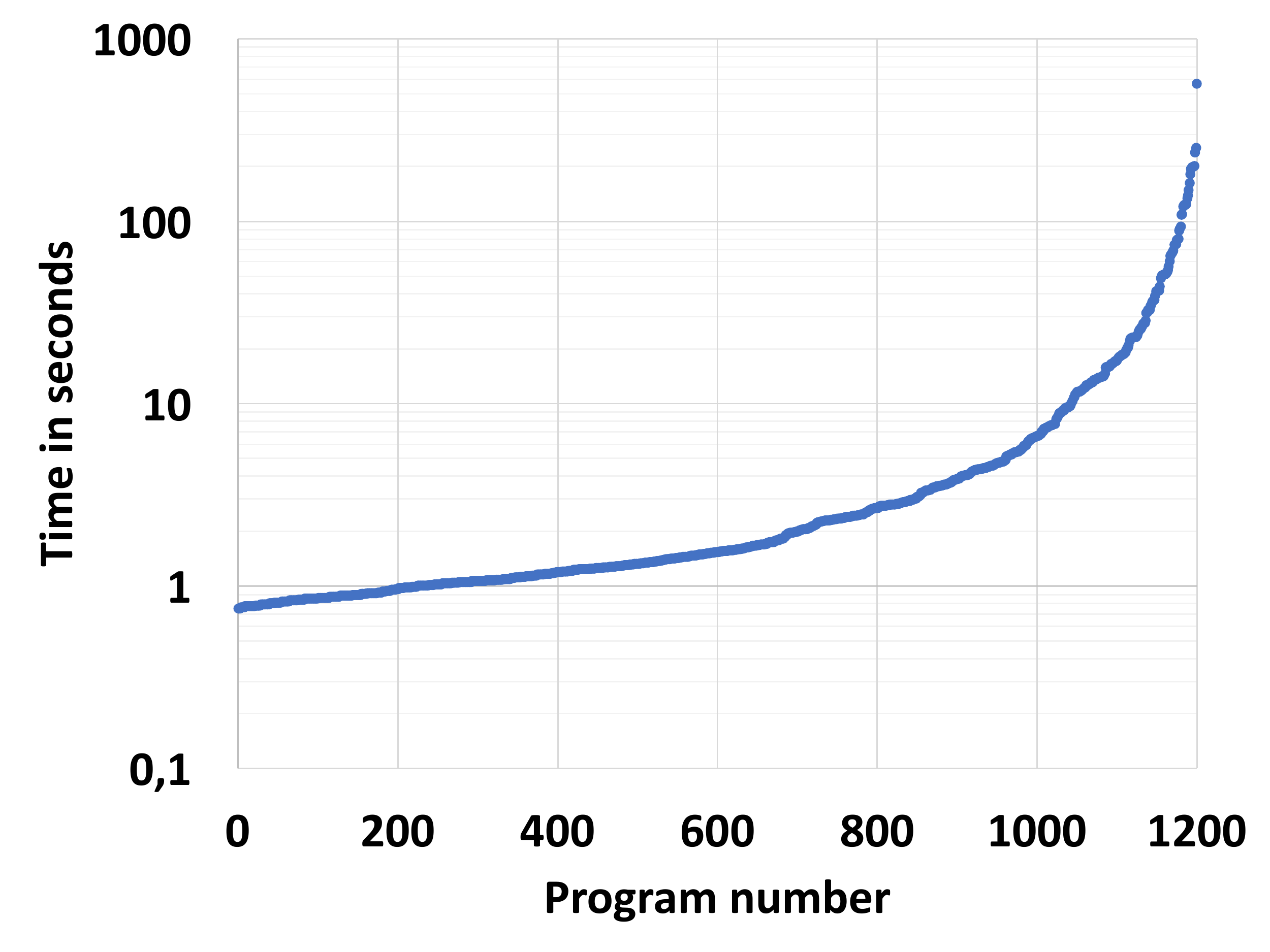}
     \end{center}
     \vspace{-18pt}
    \caption{        Generation of All Valid Executions (form 3 to 8 memory events).
     }    \label{fig:all_execs_scalability}
  \vspace{-20pt}    
\end{wrapfigure}

As described in Section~\ref{sec:implementation}, the generation of
all valid executions is based on a single AllSAT procedure.
Figure~\ref{fig:all_execs_scalability} shows a scalability
evaluation when generating all valid executions of 1200 program
instances, each with from 3 to 8 memory events (200 programs for each
configuration). The x-axis refers to the program number, ordered first by
number of memory events, and then by increasing
execution time, while the y-axis reports the execution time (in
seconds on an Intel i7-6700 @ 3.4GHz) on a logarithmic scale. The
results show that the proposed approach is able to analyze programs
with 7 memory events in fewer than 10 seconds, providing reasonable
responsiveness to deal with small, but informative, programs.

\paragraph{Behavioral Coverage Constraints}

For the coverage constraints analysis, we first extracted a subset of
the 1200 tests, considering only the ones that could produce at least 5
different outputs. There were 288 such tests.
For each test, we ran the JavaScript engine 500
times, and performed an analysis using 11 predicates, each of which
corresponds to a sub-part of the \memorymodel, as well as some
additional formulae. During this evaluation, the average
computation time required to perform the behavioral coverage constraints
analysis was 3.25 seconds, with a variance of 0.37 seconds.
 
\section{Results of the Formal Analyses}
\label{sec:results_analysis}

In this Section we provide an overview of the results of the formal
analyses for the \ES \memorymodel.

\subsubsection{Circular relations definition}
In the original \memorymodel, a subset of the relations were specified
using circular definitions. More specifically, using the
notation a $\rightarrow$ b as ``the definition of a depends on b'',
the loop was \SW $\rightarrow$ \RF $\rightarrow$ \RBF $\rightarrow$
\HB $\rightarrow$ \SW. Cyclic definitions can result in vacuous
constraints, and in the case of binary relations, this manifests as solutions with
unconstrained tuples that belong to all relations involved in the
cycle. In order to solve this problem, the definition of \RBF was
changed so that it no longer depends on \emph{Happens Before}.
In addition, the memory model was extended
with a property called \emph{Valid Coherent Reads} that constrains
the possible tuples belonging to the \RBF relation.

\subsubsection{Misalignment of the ComposeWriteEventBytes}
The memory model defines a \RBF relation, and checks whether the
tuples belonging to it are valid by relying on a function called
\emph{ComposeWriteEventBytes}. Given a list of writes, the
\emph{ComposeWriteEventBytes} function creates a vector of values
associated with a read event; however, the index for each write event
was not correct, resulting in a misalignment w.r.t. the \RBF
relation. An additional offset was added in order to fix the
problem.

\subsubsection{Distinct events quantification}
Another problem encountered while analyzing the \ES memory model was
caused by a series of inconsistent constraints. One example of
inconsistency was in the definition of the \HB relation which prescribes
that for any two events $ev_1$ and $ev_2$ with overlapping ranges, whenever
$ev_1$ is of type \emph{Init}, $ev_2$ should be of a different
type (i.e., not \emph{Init}). However, there was no constraint stating that
$ev_1$ and $ev_2$ have to be distinct, and certainly, whenever
$ev_1$ and $ev_2$ are not distinct then this expression is
unsatisfiable.

A similar inconsistency was found in the definition of
the \MO relation. In this case, if the SW relation contains
the pair $(ev_1, ev_2)$, and $(ev_1, ev_2) \in $ HB, then the MO
should contain $(ev_1, ev_2)$. However, this is inconsistent with another
constraint requiring that no event $ev_3$ should exist operating
on the same memory addresses as $ev_2$ such that both $(ev_1, ev_3) \in$ MO and
$(ev_3, ev_2) \in$ MO. This constraint is false when $ev_1 = ev_2 = ev_3$.
Both the \HB and the \MO relations initially permitted any pairs of elements to
be related (including two equal elements). The solution was 
to only allow pairs of distinct events in these relations.

The definition of the \RBF relation stated that each read or modify
event $ev_1R$ is associated with a list of pairs of byte indices and
write or modify events.  The 
definition did not specifically preclude allowing modify events to read
from themselves. This does not cause any particular issues at the
formal model level, but it is not clear what the implication
at the \JS engine implementation level would be. In order to resolve
this issue, the definition of the \RBF relation was
modified to allow only events that are distinct to be related by \RBF.

\subsubsection{Outputs coverage on \ES engines}

As described in Section~\ref{sec:formal_analyses}, the litmus test
analysis can result in three possible outcomes, e.g., $E_{x}(P) \setminus
\text{\emph{VE}}(P) \neq \emptyset$ when the engine violates the
specification, $E_{x}(P) = \text{\emph{VE}}(P)$ when the engine
matches the specification, and $E_{x}(P) \subset \text{\emph{VE}}(P)$
when the engine is more restrictive than the specification.  Typically,
such an analysis is designed to find bugs in the software
implementation of the memory model \cite{Jade10,Batty15}, focusing
on the first case ($E_{x}(P) \setminus \text{\emph{VE}}(P) \neq \emptyset$).
However, in this project, the last case was most prevalent,
where $E_{x}(P)$ is significantly smaller than $\text{\emph{VE}}(P)$.

For instance, when we ran the 288 examples with at least
5 possible outputs (from Section~\ref{sec:evaluation}) 1000 times for each combination of program and \JS
engine, the overall output coverage reached $75\%$, but for $1/6$ of the
examples, the coverage did not exceed $50\%$, and some were even below
$15\%$\footnote{\scriptsize{On an x86 machine, and with the latest
  version of the engines available on October 1st, 2017.}}.

This situation (frequently having far fewer observed behaviors than allowed
behaviors) guided our
development of alternative analyses, such as the generation of the
behavioral coverage constraints, to help developers
understand the relationship between an engine's implementation and the memory
model specification. Future improvements of \JS engines will likely
be less conservative, meaning that more behaviors will be covered.
The tests produced in this project will be essential to ensure that no bugs are
introduced. Currently, we are in the process
of adapting the litmus tests so that they can be included as part
of the official TEST262 test suite for the \ES \memorymodel.
 
\section{Conclusion}
\label{sec:conclusion}

Extending JavaScript, the language used by nearly all web-based
interfaces, to support shared memory operations warrants
the use of extensive verification techniques. In this work, we have
presented a tool that has been developed in order to support the
design and development of the \ES \memorymodel. The formal
analysis of the original specification allowed us to identify a
number of potential issues and inconsistencies. The evaluation of the
valid executions and litmus tests coverage analysis identified a
conservative level of optimization in current engine
implementations. This situation motivated us to develop a specific
technique for understanding differences between the \memorymodel
specification and \JS engine implementations.

Future extensions to this work will consider providing additional
techniques to help developers improve code optimizations in
\JS engines. Techniques such as the synthesis of equivalent
programs, and automated value instantiation given a parametric
program will provide additional analytical capabilities able to
identify possible directions for code optimization. Moreover, we
will also consider integration with other constraint solving
engines in order to deal with more complex programs.

\bibliographystyle{abbrv}

\end{document}